\documentclass[twocolumn,showpacs,eqsecnum,prb]{revtex4}
\usepackage{graphicx}%
\usepackage{dcolumn}
\usepackage{amsmath}

\makeatletter
\def\btt#1{\texttt{\@backslashchar#1}}%
\DeclareRobustCommand\bblash{\btt{\@backslashchar}}%
\makeatother

\begin{document}

\title[Short Title]{DC Josephson Effect in SNS Junctions of\\
Anisotropic Superconductors}

\author{Yasuhiro Asano}
\email{asano@eng.hokudai.ac.jp}
\affiliation{%
Department of Applied Physics, Hokkaido University, 
Sapporo 060-8628, Japan
}%

\date{\today}

\begin{abstract}
A formula for the Josephson current
between two superconductors with anisotropic pairing symmetries is derived
based on the mean-field theory of superconductivity. 
Zero-energy states formed at the junction interfaces is one of basic 
phenomena in anisotropic superconductor junctions.
In the obtained formula, effects of the 
zero-energy states on the Josephson current are taken into account through
the Andreev reflection coefficients of a quasiparticle.
In low temperature regimes, the formula can describe 
an anomaly in the Josephson current which is a direct consequence 
of the exsitence of zero-energy states. 
It is possible to apply the formula to junctions consist of 
superconductors with spin-singlet Cooper pairs
and those with spin-triplet Cooper pairs.
\end{abstract}

% PACS, 
\maketitle

\section{introduction}
The discoveries of the high-$T_c$ superconductors~\cite{bednorz} have 
stimulated an intensive research in this field. A symmetry
of a Cooper pair is an important information to understand
the mechanism of high-$T_c$ superconductivity. The Josephson effect
in anisotropic superconductors has attracted considerable interst in recent
years because high-$T_c$ superconductors may have the $d_{x^2-y^2}$-wave 
pairing symmetry~\cite{sigrist,wollman}. 
So far, transport properties in various junctions of the
$d$-wave superconductors have been discussed in a number of 
studies~\cite{kashiwaya,lofwander,yip,bruder,kuklov,barash,tanaka,samanta,riedel,golubov,fogelstrom,zhang,zhu,ohashi,matsumoto,nagato,asano}.
In anisotropic superconductors, a sign of the pair potential depends on 
a direction of a quasiparticle's motion. As a consequence, zero-energy
states (ZES's)~\cite{hu} are formed at the normal metal/ 
superconductor (NS) interface
when the potential barrier at the interface is large enough. 
The ZES's have been seen in the conductance 
spectra of tunnel junctions~\cite{wei,iguchi}.
It is known that the ZES's cause a low-temperature anomaly of 
the Josephson current in SIS junctions of the $d$-wave 
superconductor~\cite{barash,tanaka}.

 The anisotropic superconductivity itself has been an important topic in
condensed matter physics since unconventional superconductivity
was found in heavy-fermion materials 
such as, CeCu$_2$Si$_2$, UBe$_{13}$ and 
UPt$_3$~\cite{steglich,ott,stewart}. In a recent study, the anisotropic 
superconductivity was reported in a layered perovskite  
Sr$_2$RuO$_4$~\cite{maeno}.
Some of interesting effects of the anisotropy in the pairing symmetry on
Josephson current are revealed in previous 
work~\cite{geshkenbein,millis,sigrist2}. 
However, in order to study the contribution of the ZES's 
to the Josephson current,
we have to pay careful attention to a boundary condition of a
wavefunction at the junction interface~\cite{tanaka2}.
Thus an expression of the Josephson current that describes the
effects of the ZES's is desirable to study an aspect of
transport properties in anisotropic supercondutor junctions.
So far, such formular for the Josephson current is obtained in
SIS junctions of $d_{x^2-y^2}$ superconductors~\cite{tanaka}.
However there is no general fomula which can be applied to
junctions of spin-triplet superconductors.

In this paper, we derive a formula for the Josephson current
in junctions of anisotropic superconductors with 
spin-singlet and spin-triplet Cooper pairs.
The results are an extension of the Furusaki-Tsukada 
formula for $s$-wave 
superconductor junctions~\cite{furusaki}.
Effects of the ZES's on the Josephson 
current is naturally taken into account in the obtained formula 
through the Andreev reflection~\cite{andreev} coefficients (ARC's) 
of a quasiparticle.
The low-temperature anomaly in the Josephson current is described by 
the dependence of the ARC's on temperatures.
Throughout this paper, we take the units of $\hbar=k_B=1$, where $k_B$ is the
Boltzmann constant.  

This paper is organized as follows. In Sec.~II, we derive the Josephson
current formula based on the mean-field theory of superconductivity. 
In Sec.~III, the formula is applied to junctions 
of superconductors with spin-singlet and spin-triplet Copper pairs.
The conclusion is given in Sec.~IV.

\section{Josephson Current Formula I}

\begin{figure}[bp]
\includegraphics[width=8.0cm]{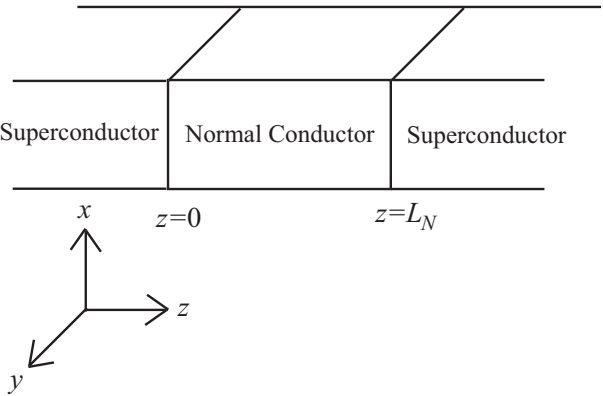}
\caption{
The SNS junction of anisotropic superconductors is illustrated. 
The phase of the pair potential on the left (right) superconductor 
is $\varphi_L$  ($\varphi_R$).
}
\label{system}
\end{figure}

Let us consider SNS junctions as shown in 
Fig.~\ref{system}, where 
the length of the normal metal is $L_N$ and the cross section of the junction
is $S_J$. 
The (BCS) Hamiltonian in the mean-field approximation reads
\begin{align}
H_{MF}=&\frac{1}{2}\int\!\! d\boldsymbol{r} \int\!\! d\boldsymbol{r}' 
\left[ {\tilde c^\dagger(\boldsymbol{r})} 
\delta(\boldsymbol{r}-\boldsymbol{r}')
 {\hat h}_0(\boldsymbol{r}') {\tilde c^{ }
(\boldsymbol{r}')} \right. \nonumber \\
 &- \tilde{c}^t(\boldsymbol{r}) \delta(\boldsymbol{r}-\boldsymbol{r}') 
{\hat h}^\ast_0
(\boldsymbol{r}') 
\left\{\tilde{c}^{\dagger}(\boldsymbol{r}')\right\}^t \nonumber \\
 &+ 
{\tilde c^\dagger(\boldsymbol{r})} 
{\hat \Delta}(\boldsymbol{r}-\boldsymbol{r}') 
\left\{{\tilde c^{\dagger}(\boldsymbol{r}')}\right\}^t \nonumber \\
 &-\left.
 {\tilde c^t(\boldsymbol{r})} 
{\hat \Delta}^\ast(\boldsymbol{r}-\boldsymbol{r}') 
{\tilde c^{ }(\boldsymbol{r}')} \right], \label{bcs2}\\
{\hat h}_0(\boldsymbol{r})=&
\left[-\frac{\nabla^2}{2m}+V_0(\boldsymbol{r})-\mu_F\right]{\hat \sigma}_0 
+ \boldsymbol{V} (\boldsymbol{r}) \cdot {\hat{\boldsymbol{\sigma}}},\\
{\tilde c(\boldsymbol{r})}\equiv& 
\left(\begin{array}{c}
c_\uparrow(\boldsymbol{r}) \\
c_\downarrow(\boldsymbol{r})
\end{array}\right), 
\label{defpsi}
\end{align}
where $c_\sigma(\boldsymbol{r})$ is the annihilation operator of an electron 
at $\boldsymbol{r}$ with spin $\sigma=\uparrow$ or $\downarrow$, 
$\left\{{\tilde c(\boldsymbol{r})}\right\}^t$ is the transpose
of Eq.~(\ref{defpsi}), $\hat{\sigma}_0$ is the unit matrix of $2\times 2$, 
and $\mu_F$ is the Fermi energy.
Spin-independent potential is represented by $V_0(\boldsymbol{r})$ 
which includes the barrier potential at the two NS interfaces given by
$V_b \left\{\delta(z) + \delta(z-L_N)\right\}$.
Spin-orbit scattering in the normal metal is denoted by 
$\boldsymbol{V}(\boldsymbol{r}) \cdot {\hat{\boldsymbol{\sigma}}}$.
A pair potential between an electron with ($\sigma$, $\boldsymbol{r}$) and  
that with ($\sigma'$, $\boldsymbol{r}'$) is described by 
$\Delta_{\sigma,\sigma'}(\boldsymbol{r}-\boldsymbol{r}')$. 
In the normal segment ($0<z<L_N$), the pair potential is taken to be zero. 
In what follows, $2\times 2$ matrices are indicated by $\widehat{\cdots}$. 
The pair potential is given by
\begin{equation}
{\hat \Delta}(\boldsymbol{r}) = 
\begin{cases} i d_0(\boldsymbol{r}) \hat{\sigma}_2 & \text{ singlet } \\
i (\boldsymbol{d}(\boldsymbol{r}) \cdot {\hat{\boldsymbol{\sigma}}} )
\hat{\sigma}_2
 & \text{ triplet },
\end{cases}
\end{equation}
where $\hat{\sigma}_j$ with $ j = 1, 2$ and 3 are the Pauli's matrices.
The pair potential satisfies a relation
\begin{equation}
-{\hat \Delta}^t(\boldsymbol{r}'-\boldsymbol{r})={\hat \Delta}(\boldsymbol{r}-\boldsymbol{r}').
\end{equation}
The Hamiltonian in Eq.(\ref{bcs2}) is diagonalized by the Bogoliubov
transformation,
\begin{align}
\left[ \begin{array}{c}
\tilde{c}(\boldsymbol{r}) \\
\{\tilde{c}^\dagger(\boldsymbol{r})\}^t 
\end{array}
\right]
=&\sum_\lambda 
\left[\begin{array}{cc}
\hat{u}_\lambda(\boldsymbol{r}) & \hat{v}_\lambda^\ast(\boldsymbol{r}) \\
\hat{v}_\lambda(\boldsymbol{r}) & \hat{u}_\lambda^\ast(\boldsymbol{r}) 
\end{array}
\right]
\left[ \begin{array}{c}
\tilde{\alpha}_\lambda \\
\{\tilde{\alpha}_\lambda^\dagger\}^t 
\end{array}
\right],\\
H_{MF}=&\sum_\lambda\tilde{\alpha}^\dagger \, \hat{E}_\lambda \, 
\tilde{\alpha}_\lambda, \\
\hat{E}_\lambda =&
\left[
\begin{array}{cc}
E_{\lambda,1} & 0 \\
0 & E_{\lambda,2}
\end{array}
\right],\\
\intertext{where} 
\tilde{\alpha}_\lambda \equiv &\left( \begin{array}{c}
\alpha_{\lambda,\uparrow} \\
\alpha_{\lambda,\downarrow} 
\end{array}
\right),
\end{align}
denotes the annihilation operator of a Bogoliubov quasiparticle.
The wavefunctions satisfy the Bogoliubov-de Gennes (BdG) 
equation~\cite{degennes},
\begin{eqnarray}
&\displaystyle{\int}& \!\!\!d\boldsymbol{r}'
 \left[
  \begin{array}{cc} 
  \delta(\boldsymbol{r}-\boldsymbol{r}') {\hat h}_0(\boldsymbol{r}') 
& \hat{\Delta}(\boldsymbol{r}-\boldsymbol{r}') \\
  -\hat{\Delta}^\ast(\boldsymbol{r}-\boldsymbol{r}') 
& -\delta(\boldsymbol{r}-\boldsymbol{r}') {\hat h}^\ast_0(\boldsymbol{r}')
  \end{array} 
 \right]\nonumber \\
&\times& \left[
 \begin{array}{c}
  \hat{u}_\lambda(\boldsymbol{r}')\\
  \hat{v}_\lambda(\boldsymbol{r}')
 \end{array}
 \right] 
= 
 \left[
 \begin{array}{c}
  \hat{u}_\lambda(\boldsymbol{r})\\
  \hat{v}_\lambda(\boldsymbol{r})
 \end{array}
 \right]\hat{E}_\lambda. \label{bdg}
\end{eqnarray}
When the wavefunction 
\begin{equation}
 \left[
 \begin{array}{c}
  \hat{u}_\lambda(\boldsymbol{r})\\
  \hat{v}_\lambda(\boldsymbol{r})
 \end{array}
 \right],
\end{equation}
is belonging to a positive eigenvalue $\hat{E}_\lambda$, the wavefunction
\begin{equation}
 \left[
 \begin{array}{c}
  \hat{v}_\lambda^\ast(\boldsymbol{r})\\
  \hat{u}_\lambda^\ast(\boldsymbol{r})
 \end{array}
 \right],
\end{equation}
is belonging to $-\hat{E}_\lambda$. They satisfy the 
following relations
\begin{align}
\int d\boldsymbol{r}\left\{
\hat{u}_\lambda^\dagger(\boldsymbol{r}) \hat{u}_{\lambda'}(\boldsymbol{r})
+ \hat{v}_\lambda^\dagger(\boldsymbol{r}) \hat{v}_{\lambda'}(\boldsymbol{r})
\right\}
=&\delta_{\lambda,\lambda'}\hat{\sigma}_0, \\
\int d\boldsymbol{r}\left\{
\hat{u}_\lambda^\dagger(\boldsymbol{r}) \hat{v}_{\lambda'}^\ast(\boldsymbol{r})
+ \hat{v}_\lambda^\dagger(\boldsymbol{r}) \hat{u}_{\lambda'}^\ast(\boldsymbol{r})
\right\}
=&\hat{0}, \\
{\sum_\lambda}' \left\{
\hat{u}_\lambda(\boldsymbol{r}) \hat{u}_{\lambda}^\dagger(\boldsymbol{r}')
+\hat{v}_\lambda^\ast(\boldsymbol{r}) \hat{v}_{\lambda}^t(\boldsymbol{r}')
\right\}=& \delta(\boldsymbol{r}-\boldsymbol{r}')
 \hat{\sigma}_0, \\
{\sum_\lambda}' \left\{
\hat{u}_\lambda(\boldsymbol{r}) \hat{v}_{\lambda}^\dagger(\boldsymbol{r}')
+\hat{v}_\lambda^\ast(\boldsymbol{r}) \hat{u}_{\lambda}^t(\boldsymbol{r}')
\right\}=&\hat{0}, 
\end{align}
where ${\sum_\lambda}'$ is a summation over $\lambda$ with positive 
eigenvalues.
The local charge density is defined by
\begin{equation}
P(\boldsymbol{r},\tilde{t}) = -e\: {\tilde c}^\dagger(\boldsymbol{r},\tilde{t})
\: {\tilde c}(\boldsymbol{r},\tilde{t}),
\label{cdensity}
\end{equation}
where $\tilde{t}$ is a time. 
The current conservation low implies,
\begin{equation}
\frac{\partial}{\partial \tilde{t}}P(\boldsymbol{r},\tilde{t}) 
+ \nabla \cdot \boldsymbol{J}(\boldsymbol{r},\tilde{t})=0.\label{conserve}
\end{equation}
The Josephson current between the two superconductors
is calculated from the expectation value of Eq.~(\ref{conserve})
\begin{widetext}
\begin{align}
\boldsymbol{J}(\boldsymbol{r})=& 
\displaystyle{\frac{e}{4mi}}\lim_{\boldsymbol{r}'\rightarrow \boldsymbol{r}}
\left( \nabla_{\boldsymbol{r}'}-\nabla_{\boldsymbol{r}} \right) 
T \sum_{\omega_n} 
\textrm{ Tr} \; \check{\cal G}_{\omega_n}(\boldsymbol{r},\boldsymbol{r}'),\label{j1}\\
\check{\cal G}_{\omega_n}(\boldsymbol{r},\boldsymbol{r}') =& 
{\sum_{\lambda}}' 
\left[
\left[\begin{array}{c}
\hat{u}_\lambda(\boldsymbol{r}) \\
\hat{v}_\lambda(\boldsymbol{r})\end{array}
\right]
\left[ i\omega_n \hat{\sigma}_0 - \hat{E}_\lambda\right]^{-1}
\left[\begin{array}{c}
\hat{u}_\lambda(\boldsymbol{r}')\\ \hat{v}_\lambda(\boldsymbol{r}')
\end{array}\right]^\dagger 
+ \left[\begin{array}{c}
\hat{v}_\lambda^\ast(\boldsymbol{r}) \\
\hat{u}_\lambda^\ast(\boldsymbol{r})\end{array}
\right]
\left[ i\omega_n \hat{\sigma}_0 + \hat{E}_\lambda\right]^{-1}
\left[\begin{array}{c} \hat{v}_\lambda^\ast(\boldsymbol{r}')\\
 \hat{u}_\lambda^\ast(\boldsymbol{r}')\end{array}\right]^\dagger 
\right],
\end{align}
where $T$ is a temperature,  
$\check{\cal G}_{\omega_n}(\boldsymbol{r},\boldsymbol{r}')$ 
is the Matsubara Green function of the SNS junctions and
$\check{\cdots}$ indicates $4\times 4$ matrices.
On the derivation of Eq.~(\ref{j1}), 
we have assumed that the amplitude of the pair potential
is much smaller than the Fermi energy $\mu_F$.

In the superconductors, we assume that the all potentials are uniform. 
Thus the BdG equation in Eq.~(\ref{bdg}) is
given in the Fourier representation,
\begin{equation}
 \left[
  \begin{array}{cc} 
  \xi_{\boldsymbol{k}} \hat{\sigma}_0 
  & \hat{\Delta}(\boldsymbol{k}) \\
  -\hat{\Delta}^\ast(-\boldsymbol{k}) 
 & - \xi_{\boldsymbol{k}}\hat{\sigma}_0
  \end{array}
 \right]
 \left[
 \begin{array}{c}
  \hat{u}_{\boldsymbol{k}}\\
  \hat{v}_{\boldsymbol{k}}
 \end{array}
 \right] 
= 
 \left[
 \begin{array}{c}
  \hat{u}_{\boldsymbol{k}}\\
  \hat{v}_{\boldsymbol{k}}
 \end{array}
 \right] \hat{E}_{\boldsymbol{k}}, 
\end{equation}
where $\xi_{\boldsymbol{k}}=\boldsymbol{k}^2/(2m)-\mu_F$ and 
\begin{align}
\hat{\Delta}(\boldsymbol{r}-\boldsymbol{r}') =& \sum_{\boldsymbol{k}}
\hat{\Delta}(\boldsymbol{k})
\textrm{ e}^{i\boldsymbol{k}\cdot(\boldsymbol{r}-\boldsymbol{r}')}, \\
\hat{\Delta}(\boldsymbol{k}) =& \begin{cases}
i d_0(\boldsymbol{k}) \hat{\sigma}_2  & \text{singlet} \\
i (\boldsymbol{d}(\boldsymbol{k}) \cdot \hat{\boldsymbol{\sigma}}) 
\hat{\sigma}_2 &
\textrm{triplet} \end{cases}\label{deltak}.
\end{align}
Since relations
\begin{align}
d_0(-\boldsymbol{k}) =& d_0(\boldsymbol{k}), \\
\boldsymbol{d}(-\boldsymbol{k}) =& - 
\boldsymbol{d}(\boldsymbol{k}),\label{invp}
\end{align}
are satisfied in the momentum space, one finds
\begin{equation}
-\hat{\Delta}^t(-\boldsymbol{k}) = \hat{\Delta}(\boldsymbol{k}).
\end{equation}
When  $z<z'\leq 0$, the Green function can be calculated as
\begin{eqnarray}
& &\check{\cal G}_{\omega_n}(\boldsymbol{r},\boldsymbol{r}')= -i m\omega_n \sum_{\boldsymbol{p}}
\chi_{\boldsymbol{p}}(\boldsymbol{\rho})\; 
\chi_{\boldsymbol{p}}^\ast(\boldsymbol{\rho}')
\check{\Phi}_L \nonumber \\
&\times &\left[ \left\{ 
\left(\begin{array}{c} \hat{u}_+^e \\ \hat{v}_+^e \end{array}\right)
\hat{K}_p(k^e_+,z) +
\left(\begin{array}{c} \hat{u}_+^h \\ \hat{v}_+^h \end{array}\right)
\hat{K}_p(k^h_+,z) \hat{a}_1  
+ \left(\begin{array}{c} \hat{u}_-^e \\ \hat{v}_-^e \end{array}\right)
\hat{K}_p(-k^e_-,z) \hat{b}_1
\right\} 
 \hat{K}_p(-k^e_+,z')
\left(\begin{array}{cc} k^e_{1,+} & 0 \\ 0 & k^e_{2,+} 
\end{array}\right)^{-1}
\hat{\Omega}_+^{-1}
\left(\begin{array}{c} \hat{u}_+^e \\ \hat{v}_+^e \end{array}\right)^\dagger
\right. \nonumber\\
&+&
\left\{ \left(\begin{array}{c} \hat{u}_-^h \\ \hat{v}_-^h \end{array}\right)
\hat{K}_p(-k^h_-,z) + 
\left(\begin{array}{c} \hat{u}_-^e \\ \hat{v}_-^e \end{array}\right)
\hat{K}_p(-k^e_-,z) \hat{a}_2 
\left(\begin{array}{c} \hat{u}_+^h \\ \hat{v}_+^h \end{array}\right)
\hat{K}_p(k^h_+,z) \hat{b}_2 \right\} 
\left. \hat{K}_p(k^h_-,z')
\left(\begin{array}{cc} k^h_{1,-} & 0 \\ 0 & k^h_{2,-} \end{array}\right)^{-1}
\hat{\Omega}_-^{-1}
\left(\begin{array}{c} \hat{u}_-^h \\ \hat{v}_-^h \end{array}\right)^\dagger
\right] \nonumber\\
&\times& \check{\Phi}_L^\ast, \label{g1}
\end{eqnarray}
\end{widetext}
with 
\begin{align}
k_{l,\pm}^{e}=&\sqrt{2m(\mu_F-\epsilon(\boldsymbol{p})+i\Omega_{l,\pm})},
\label{kepm}\\
k_{l,\pm}^{h}=&\sqrt{2m(\mu_F-\epsilon(\boldsymbol{p})-i\Omega_{l,\pm})},
\label{khpm}\\
\Omega_{l,\pm}=& \sqrt{\omega_n^2+|\Delta_{l,\pm}|^2},\\
\epsilon(\boldsymbol{p}) =& \frac{\boldsymbol{p}^2}{2m},\\
d_{0,\pm} =& d_0(\boldsymbol{p},\pm k_z) \label{d0pm},\\
\boldsymbol{d}_\pm =& \boldsymbol{d}(\boldsymbol{p}, \pm k_z) \label{dpm},\\ 
\hat{\Delta}_\pm =& \begin{cases}
i d_{0,\pm} \hat{\sigma}_2 & \text{singlet}\\
i (\boldsymbol{d}_\pm \cdot \hat{\boldsymbol{\sigma}}) \hat{\sigma}_2 
& \text{triplet}
\end{cases},
\end{align}
\begin{align}
\boldsymbol{q}_\pm =& i \boldsymbol{d}_\pm \times \boldsymbol{d}_\pm^\ast, \\
\hat{K}_p(k,z)=&
\left(\begin{array}{cc}\textrm{ e}^{ik_1 z} & 0 \\ 0 & \textrm{ e}^{ik_2 z}
\end{array}\right), \\
\hat{\Omega}_\pm =&\left(\begin{array}{cc}
\Omega_{1,\pm} & 0 \\
0 & \Omega_{2,\pm} \end{array}\right),\\
\check{\Phi}_j=&
\left( \begin{array}{cc} \textrm{ e}^{i\varphi_j}\hat{\sigma}_0 & 0\\
0& \textrm{ e}^{-i\varphi_j}\hat{\sigma}_0 \end{array}\right),\\
\chi_{\boldsymbol{p}}(\boldsymbol{\rho})=&
\frac{\exp(i\boldsymbol{p}\cdot \boldsymbol{\rho})}{\sqrt{S_J}},
\end{align}
where $\varphi_j$ for $j=L$ or $R$ is the phase of the superconductor,
 $\boldsymbol{p}=(k_x,k_y)$ and $\boldsymbol{\rho}= (x,y)$.
The amplitude of the pair potential for unitary states is defined by
\begin{equation}
|\Delta_{l,\pm}|=|\Delta_{\pm}|
= \begin{cases} 
|d_{0,\pm}| & \text{singlet}\\
|\boldsymbol{d}_{\pm}| & \text{triplet}
\end{cases}.
\end{equation}
In unitary states, these amplitudes are 
independent of $l$, where
$l$ indicates the spin configuration of a quasiparticle. 
The amplitude of the pair potential depends on the spin configuration
of a quasiparticle in nonunitary states,
\begin{equation}
|\Delta_{l,\pm}|=\begin{cases}
\sqrt{|\boldsymbol{d}_\pm|^2+ |\boldsymbol{q}_\pm|}  & (l=1) \\ 
\sqrt{|\boldsymbol{d}_\pm|^2- |\boldsymbol{q}_\pm|}  & (l=2)  
\end{cases}.
\end{equation}
In Eqs.~(\ref{kepm}) and (\ref{khpm}), 
$k^{e(h)}_{l,\pm}$ is the wavenumber in the electron (hole) 
branch for $l$-th spin state. In the following, we approximately
describe these wavenumbers as
$k^{e(h)}_{l,\pm}\approx k_z = \sqrt{2m(\mu_F-\epsilon(\boldsymbol{p}))}$
as shown in Eqs.(\ref{d0pm}) and (\ref{dpm}), where 
$(\boldsymbol{p},\pm k_z)$ is the wavenumber on the 
Fermi surface. The $l$-th column of 
\begin{equation}
\left( \begin{array}{c}
\hat{u}^{e(h)}_\pm \\
\hat{v}^{e(h)}_\pm 
\end{array}
\right),
\end{equation}
corresponds to the wavefunction of $l$-th spin state in the electron (hole) 
branch.
The reflection coefficients from the left superconductor to the 
left superconductor are defined in a matrix form~\cite{furusaki}
\begin{align}
\hat{a}_{j=1,2}=&\left(\begin{array}{cc}
a_j(1,1) & a_j(1,2)\\
a_j(2,1) & a_j(2,2)
\end{array}\right), \\
\hat{b}_{j=1,2}=&\left(\begin{array}{cc}
b_j(1,1) & b_j(1,2)\\
b_j(2,1) & b_j(2,2)
\end{array}\right).
\end{align}
The ARC from the $l$-th spin state in the
electron (hole) branch to the $l'$-th spin state in the hole (electron) 
branch is denoted by $\hat{a}_1(l',l)$, ($\hat{a}_2(l',l)$). 
In the same way, $\hat{b}_1(l',l)$ ($\hat{b}_2(l',l)$) is the normal reflection
coefficient from the $l$-th spin state in the electron (hole) branch
to the $l'$-th spin state in the electron (hole) branch. 
These reflection coefficients depend on  
$\boldsymbol{p}$ which indicates
the propagating channel at the left NS interface.
Substituting Eq.~(\ref{g1}) into Eq.~(\ref{j1}), the Josephson
current becomes 
\begin{align}
J=&\frac{ie}{2} \sum_{\boldsymbol{p}} 
T\sum_{\omega_n} \textrm{ Tr} \;\; \omega_n \nonumber \\
\times& \left[
\left(\begin{array}{c} \hat{u}^{h}_+ \\ \hat{v}^{h}_+ \end{array}\right)
\hat{a}_1 \hat{\Omega}_+^{-1}
\left(\begin{array}{c}
\hat{u}^{e}_+ \\
 \hat{v}^{e}_+ \end{array}\right)^\dagger \right.\nonumber \\
 &-\left.\left(\begin{array}{c}
 \hat{u}^{e}_- \\ \hat{v}^{e}_- \end{array}\right)
\hat{a}_2 \hat{\Omega}_-^{-1}
\left(\begin{array}{c} \hat{u}^{h}_- \\ \hat{v}^{h}_- \end{array}
\right)^\dagger 
 \right].\label{j2}
\end{align}
The expression of the Josephson current in Eq.~(\ref{j2})~\cite{nishida} 
is an extension 
of the Furusaki-Tsukada formula~\cite{furusaki} for $s$-wave junctions. 

Throughout this paper, we use a representation
\begin{align}
\left( \begin{array}{c}
\hat{u}^{e}_\pm \\
\hat{v}^{e}_\pm 
\end{array}
\right) =& 
\left( \begin{array}{c}
{u}_{\pm} \hat{\sigma}_0\\
{v}_{\pm} \displaystyle{ \frac{\hat{\Delta}_\pm^\dagger}{|\Delta_\pm|}}
\end{array}
\right),\\
\left( \begin{array}{c}
\hat{u}^{h}_\pm \\
\hat{v}^{h}_\pm 
\end{array}
\right) =& 
\left( \begin{array}{c}
{v}_{\pm} \displaystyle{\frac{\hat{\Delta}_\pm}{ |\Delta_\pm|}}\\
{u}_{\pm} \hat{\sigma}_0
\end{array}
\right),\\
u_\pm =& \sqrt{\frac{1}{2}\left(1+\frac{\Omega_\pm}{\omega_n}\right)},\\
v_\pm =& \sqrt{\frac{1}{2}\left(1-\frac{\Omega_\pm}{\omega_n}\right)},
\end{align}
for unitary states. In unitary states, $\Delta_{l,\pm}$ 
is independent of $l$ because of $\boldsymbol{q}=0$.  
For nonunitary states~\cite{sigrist2}, we use 
\begin{align}
\left( \begin{array}{c}
\hat{u}^{e}_\pm \\
\hat{v}^{e}_\pm 
\end{array}
\right) =& \left( \begin{array}{c} \hat{u}_\pm \\ 
\hat{\Delta}_\pm^\dagger \hat{v}_\pm \end{array}\right)
, \\
\left( \begin{array}{c}
\hat{u}^{h}_\pm \\
\hat{v}^{h}_\pm 
\end{array}
\right) =&  \left(\begin{array}{c}
\hat{\Delta}_\pm \hat{v}_\pm^\ast \\
\hat{u}_\pm^\ast 
\end{array}\right),
\end{align}
\begin{align}
\hat{u}_\pm =& Q_\pm 
\displaystyle{
\sum_{l=1}^2 u_{l,\pm} \hat{S}_{l,\pm} }, \\
\hat{v}_\pm =&Q_\pm
\displaystyle{\sum_{l=1}^2 v_{l,\pm} 
\frac{\hat{S}_{l,\pm}}{|\Delta_{l,\pm}|}},\\ 
u_{l,\pm} =& \sqrt{\frac{1}{2}
\left(1+\frac{\Omega_{l,\pm}}{\omega_n}\right)},\\
v_{l,\pm} =& \sqrt{\frac{1}{2}
\left(1-\frac{\Omega_{l,\pm}}{\omega_n}\right)},\\
(Q_\pm)^{-2} =& 8 |\boldsymbol{q}_\pm|(|\boldsymbol{q}_\pm|+ q_{3,\pm}), \\
\hat{S}_{l,\pm} =& {\hat P}_{l,\pm} \cdot {\hat t}_l, \\
\hat{P}_{1,\pm} =& |\boldsymbol{q}_\pm|\hat{\sigma}_0 + \boldsymbol{q}_\pm 
\cdot
\hat{\boldsymbol{\sigma}}, \\
\hat{P}_{2,\pm} =& |\boldsymbol{q}_\pm|\hat{\sigma}_0 - \boldsymbol{q}_\pm 
\cdot
\hat{\boldsymbol{\sigma}}, \\
\hat{t}_1 =& \hat{\sigma}_0 + \hat{\sigma}_3, \\
\hat{t}_2 =& \hat{\sigma}_0 - \hat{\sigma}_3. 
\end{align}

\begin{figure}[h]
\includegraphics[width=8.0cm]{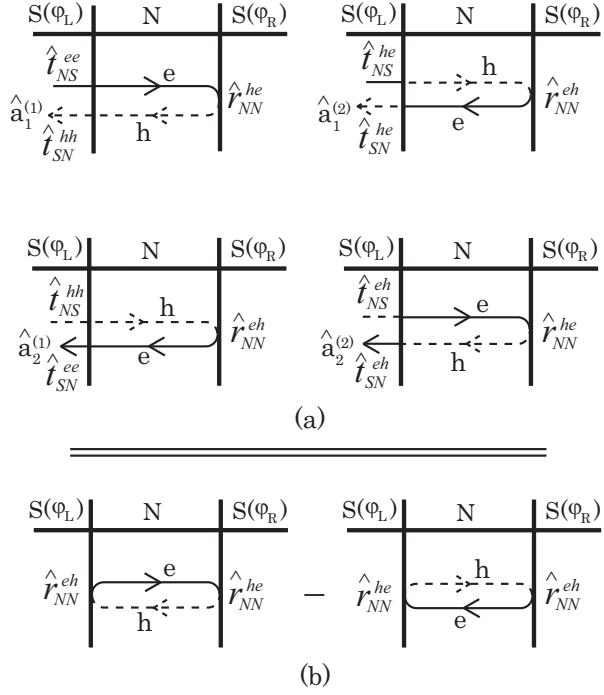}
\caption{
Four reflection processes in (a) contribute to the 
Josephson current.   
The Josephson current calculated from the four 
reflection processes in (a) is summarized in the reflection
processes in (b).
}
\label{process}
\end{figure}

In this paper, we consider four reflection processes to calculate 
$\hat{a}_1$ and $ \hat{a}_2$ as shown in Fig.~\ref{process} (a) 
and neglect all higher-order terms.
This approximation is justified
when the potential barrier
at the NS interfaces is large enough and
the transmission probability in the normal segment is
low enough. Thus in the normal segment, insulators or dirty normal 
metals are assumed.
In order to estimate $\hat{a}_1$ and $ \hat{a}_2$, 
we calculate the transmission
and the reflection coefficients at the single NS interface for 
fixed $\boldsymbol{p}$ as shown in Appendix A. 
The ARC's in Fig.~\ref{process} (a) are 
given by
\begin{widetext}
\begin{align}
\hat{a}_1^{(1)}(\boldsymbol{p}) =&  \sum_{\boldsymbol{p}'} 
\hat{t}_{SN}^{hh}(\boldsymbol{p},L) \cdot 
\hat{t}^h_{\boldsymbol{p},\boldsymbol{p}'} \cdot
\hat{r}_{NN}^{he}(\boldsymbol{p}',R) \cdot \hat{t}^e_{\boldsymbol{p}',\boldsymbol{p}}
\cdot  \hat{t}_{NS}^{ee}(\boldsymbol{p},L), \label{a11}\\
\hat{a}_1^{(2)}(\boldsymbol{p}) =&  \sum_{\boldsymbol{p}'} \hat{t}_{SN}^{he}(\boldsymbol{p},L) 
\cdot \hat{t}^e_{\boldsymbol{p},\boldsymbol{p}'}\cdot   
\;  \hat{r}_{NN}^{eh}(\boldsymbol{p}',R) \cdot \hat{t}^h_{\boldsymbol{p}', \boldsymbol{p}}
\cdot  \hat{t}_{NS}^{he}(\boldsymbol{p},L), \label{a12}\\
\hat{a}_2^{(1)}(\boldsymbol{p}) =&  \sum_{\boldsymbol{p}'} \hat{t}_{SN}^{ee}(\boldsymbol{p},L) 
\cdot \hat{t}^e_{\boldsymbol{p}, \boldsymbol{p}} \cdot
\;  \hat{r}_{NN}^{eh}(\boldsymbol{p}',R) \cdot \hat{t}^h_{\boldsymbol{p}', \boldsymbol{p}}
\cdot  \hat{t}_{NS}^{hh}(\boldsymbol{p},L),\label{a21} \\
\hat{a}_2^{(2)}(\boldsymbol{p}) =&  \sum_{\boldsymbol{p}'} \hat{t}_{SN}^{eh}(\boldsymbol{p},L) 
\cdot \hat{t}^h_{\boldsymbol{p}, \boldsymbol{p}'}  
\cdot  \hat{r}_{NN}^{he}(\boldsymbol{p}',R) \cdot \; \hat{t}^e_{\boldsymbol{p}', \boldsymbol{p}}
\cdot  \hat{t}_{NS}^{eh}(\boldsymbol{p},L), \label{a22}
\end{align}
where $\hat{t}^{e(h)}_{\boldsymbol{p}', \boldsymbol{p}}$ is the transmission 
coefficient 
of the electronlike (holelike) quasiparticle in the normal conductor,
and $\boldsymbol{p}'$ indicates the propagating channel at the right
NS interface.
The transmission coefficients in the normal metal are described 
by 
\begin{align}
{\hat t}^{e}_{\boldsymbol{p}', \boldsymbol{p}} =& i v_{\boldsymbol{p}}
 \textrm{ e}^{- i k_z' L_N}
 \int d{\boldsymbol{\rho}} \int d{\boldsymbol{\rho}}'
\, {\hat {\cal G}}_{\omega_n}^{N,e}
({\boldsymbol{\rho}}',L_N ; {\boldsymbol{\rho}},0) 
 \chi_{\boldsymbol{p}'}^\ast({\boldsymbol{\rho}}') 
\chi_{\boldsymbol{p}}({\boldsymbol{\rho}}),\label{te} \\
{\hat t}^{h}_{\boldsymbol{p}, \boldsymbol{p}'} =& i v_{\boldsymbol{p}'} 
 \textrm{ e}^{i k_z' L_N}
 \int d{\boldsymbol{\rho}} \int d{\boldsymbol{\rho}}'
\, {\hat {\cal G}}_{\omega_n}^{N,h}
(\boldsymbol{\rho},0 ; \boldsymbol{\rho}',L_N) 
 \chi_{\boldsymbol{p}}^\ast(\boldsymbol{\rho})
 \chi_{\boldsymbol{p}'}(\boldsymbol{\rho}'),\label{th} 
\end{align}
\end{widetext}

where ${\hat {\cal G}}_{\omega_n}^{N,e(h)} 
(\boldsymbol{r},\boldsymbol{r}')$ is the Green function
of the normal conductor in the electron (hole) branch~\cite{stone}.
The velocity of a quasiparicle in the $z$ direction is
$v_{\textbf{p}}$ for the propagating channel with $\boldsymbol{p}$. 
We assume that the NS interface is sufficiently clean
so that $\boldsymbol{p}$ and $\boldsymbol{p}'$ are conserved while the 
transmission and the reflection at the interfaces. 
In $\hat{a}_1^{(1)}$ in Eq.~(\ref{a11}), 
a quasiparticle-wave is initially incident into the normal segment
from the left superconductor through the channel specified by $\boldsymbol{p}$.
After the Andreev reflection at the right NS interface, 
we assume that the reflected wave  
transmits to the left superconductor through the initial channel 
of $\boldsymbol{p}$. This is because a quasiparicle in the normal 
segment has the retro property
under the time reversal symmetry~\cite{bennaker}.
The two ARC's in Eq.(\ref{j2}) are 
given by $\hat{a}_1=\hat{a}_1^{(1)}+ \hat{a}_1^{(2)}$ and 
$\hat{a}_2=\hat{a}_2^{(1)}+ \hat{a}_2^{(2)}$. 

By using Eqs.~(\ref{j2}) and (\ref{a11})-(\ref{a22}), we can derive 
a general expression of the Josephson current
\begin{align}
J=&ie\sum_{\boldsymbol{p}} \sum_{\boldsymbol{p}'} 
T\sum_{\omega_n} \textrm{ Tr} \nonumber \\
\times& 
\left[ \hat{r}_{NN}^{eh}(\boldsymbol{p},L) \cdot \hat{t}^h_{\boldsymbol{p}, \boldsymbol{p}'}
\cdot \hat{r}_{NN}^{he}(\boldsymbol{p}',R) \cdot \hat{t}^e_{\boldsymbol{p}', \boldsymbol{p}}
\right. \nonumber \\
-&\left. \hat{r}_{NN}^{he}(\boldsymbol{p},L) \cdot \hat{t}^e_{\boldsymbol{p}, \boldsymbol{p}'}
\cdot \hat{r}_{NN}^{eh}(\boldsymbol{p}',R) \cdot
\hat{t}^h_{\boldsymbol{p}', \boldsymbol{p}} \right]\label{jfin0}.
\end{align}
The reflection processes in 
Eq.~(\ref{jfin0}) are summarized in Fig.~\ref{process} (b).
Since the relations
\begin{align}
\hat{t}^e_{-\boldsymbol{p}, -\boldsymbol{p}'} =& 
\left[ \hat{t}^h_{\boldsymbol{p}, \boldsymbol{p}'}\right]^\ast, 
\label{tn1}\\
\hat{t}^h_{-\boldsymbol{p}', -\boldsymbol{p}} =& 
\left[ \hat{t}^e_{\boldsymbol{p}', \boldsymbol{p}}\right]^\ast, 
\label{tn2}\\
\hat{r}_{NN}^{eh}(-\boldsymbol{p},R) =& 
\left\{\hat{r}_{NN}^{he}(\boldsymbol{p},R)\right\}^\ast, \label{arcr}\\
\hat{r}_{NN}^{he}(-\boldsymbol{p},L) =& 
\left\{\hat{r}_{NN}^{eh}(\boldsymbol{p},L)\right\}^\ast,\label{arcl} 
\end{align}
are satisfied (see Appendices A and B), the Josephson current results in
\begin{align}
J=&-2e\textrm{ Im} 
\sum_{\boldsymbol{p}} \sum_{\boldsymbol{p}'} 
T\sum_{\omega_n} \textrm{ Tr} \nonumber \\
\times& 
\left[ \hat{r}_{NN}^{eh}(\boldsymbol{p},L) \cdot \hat{t}^h_{\boldsymbol{p}, 
\boldsymbol{p}'}
\cdot \hat{r}_{NN}^{he}(\boldsymbol{p}',R) \cdot \hat{t}^e_{\boldsymbol{p}', 
\boldsymbol{p}}
\right]. \label{jfin1}
\end{align}
The formula in Eq.~(\ref{jfin1}) can be applied to various 
Josephson junctions. For instance, it is possible to calculate 
the Josephson current in clean SIS junctions by using a relation   
$\hat{t}^{e(h)}_{\boldsymbol{p}, \boldsymbol{p}'}\propto
\delta_{ \boldsymbol{p}, \boldsymbol{p}'}
{\hat \sigma}_0$. 
We also note that the two superconductors are not 
necessary to be identical to each other.

\section{Josephson current formula II}

In this section, 
we show the ARC's of the superconductors in spin-singlet, spin-triplet 
unitary and spin-triplet 
nonunitary states because the Josephson current is described by the 
ARC's at the NS interfaces in Eq.~(\ref{jfin1}).

Firstly, we consider the superconductor with the spin-singlet Copper pairs.
The ARC's are given by
\begin{align}
{\hat r}_{NN}^{eh}(\boldsymbol{p},L)=& -i {\hat \Gamma}_{su}(\boldsymbol{p},L)
\textrm{ e}^{i\varphi_L},\\
{\hat r}_{NN}^{he}(\boldsymbol{p},R)=& -i {\hat \Gamma}_{su}^\dagger(\boldsymbol{p},R)
\textrm{ e}^{-i\varphi_R},\\
{\hat \Gamma}_{su}(\boldsymbol{p},j)=&i \Gamma_{su}(\boldsymbol{p},j){\hat \sigma}_2,\\
\Gamma_{su}(\boldsymbol{p},j)=&\left.\frac{{\bar k}_z^2 K_+ d_{0,-}}{\Xi_{su}}
\right|_j,\\
\Xi_{su}=& (H^2 + {\bar k}_z^2 )d_{0,+}d_{0,-} + H^2 K_+K_-,\label{xisu}\\ 
K_\pm =& \Omega_\pm -|\omega_n|,\\
{\bar k}_z=&k_z/k_F 
\end{align}
where $H=mV_b/k_F$ represents the strength of the potential barrier
at the NS interface and $j=L$ or $R$ symbolically denote the
character of the superconductors such as symmetries of the pair potential 
and orientation angles.

Secondly, the ARC's in spin-triplet unitary states are given by
\begin{align}
{\hat r}_{NN}^{eh}(\boldsymbol{p},L)=&- i {\hat \Gamma}_{tu}(\boldsymbol{p},L)
\textrm{ e}^{i\varphi_L},\\
{\hat r}_{NN}^{he}(\boldsymbol{p},R)=&-i{\hat \Gamma}_{tu}^\dagger
(\boldsymbol{p},R)
\textrm{ e}^{-i\varphi_R},\\
{\hat \Gamma}_{tu}(\boldsymbol{p},j)=&i 
{\boldsymbol{\Gamma}}_{tu}(\boldsymbol{p},j)
\cdot {\hat {\boldsymbol{\sigma}}}\; {\hat \sigma}_2,\\
{\boldsymbol{\Gamma}}_{tu}(\boldsymbol{p},j)=&{\bar k}_z^2 K_+ \nonumber\\
 &\!\!\!\!\!\!\!\!\!\!\!\!\!\!\!\!\!\!\!\times
\left.
\frac{ \Xi_{tu}\boldsymbol{d}_- -(H^2+{\bar k}_z^2)(\boldsymbol{d}_+^\ast \times \boldsymbol{d}_-)
\times \boldsymbol{d}_-}{\Xi_{tu}^2-\boldsymbol{D}_{tu}\cdot \boldsymbol{D}_{tu}}\right|_j,
\label{gammatu}\\
\Xi_{tu} =& (H^2+{\bar k}_z^2) 
\boldsymbol{d}_+^\ast\!\! \cdot \boldsymbol{d}_-\!\! + H^2 K_+K_-,
\label{xitu}\\
\boldsymbol{D}_{tu} =& -i (H^2+{\bar k}_z^2) 
(\boldsymbol{d}_+^\ast \times \boldsymbol{d}_-).
\end{align}
In unitary states, $\boldsymbol{d}_\pm$ often has a single component.
In such case, one finds
\begin{equation}
{\boldsymbol{\Gamma}}_{tu}(\boldsymbol{p},j)={\bar k}_z^2 K_+
\left.\frac{\boldsymbol{d}_-}{\Xi_{tu}}\right|_j,
\end{equation} 
because of $\boldsymbol{d}_+^\ast \times \boldsymbol{d}_-=0$. 

Finally we show the ARC's in nonunitary
states,
\begin{align}
{\hat r}_{NN}^{eh}(\boldsymbol{p},L)=& -i 
{\hat \Gamma}_{nu}(\boldsymbol{p},L)\textrm{ e}^{i\varphi_L},\\
{\hat r}_{NN}^{eh}(\boldsymbol{p},R)=& -i
{\hat \Gamma}_{nu}^\dagger(\boldsymbol{p},R)\textrm{ e}^{-i\varphi_R}, \\
{\hat \Gamma}_{nu}=& i{\boldsymbol{\Gamma}}_{nu}
(\boldsymbol{p},j) \cdot {\hat {\boldsymbol{\sigma}}}\; {\hat \sigma}_2,\\
{\boldsymbol{\Gamma}}_{nu}(\boldsymbol{p},j)=&{\bar k}_z^2 
\left.\frac{\boldsymbol{D}_{nu}}{\boldsymbol{D}_{nu}
\cdot \boldsymbol{D}_{nu}}\right|_j,
\label{gammanu}\\
\boldsymbol{D}_{nu}=&(H^2+{\bar k}_z^2)
\left\{ \frac{1}{2}\sum_l\frac{\boldsymbol{D}_{l,+}}
{K_{l,+}}\right\}\nonumber\\
+ &H^2
\left\{ \frac{1}{2}\sum_l\frac{K_{l,-}\boldsymbol{D}_{l,+}}{|\Delta_{l,-}|^2}
\right\},\\
\boldsymbol{D}_{1,\pm}=&\boldsymbol{d}_\pm^\ast +
i\frac{\boldsymbol{d}_\pm^\ast \times\boldsymbol{q}_\pm }
{|\boldsymbol{q}_\pm|}, \\
\boldsymbol{D}_{2,\pm}=&\boldsymbol{d}_\pm^\ast 
-i\frac{ \boldsymbol{d}_\pm^\ast\times\boldsymbol{q}_\pm }
{|\boldsymbol{q}_\pm|}, \\
K_{l,\pm}=&\Omega_{l,\pm}-|\omega_n|.
\end{align}
Detail of the calculation is shown in Appendix A, where
we derive the ARC's of the superconductors in nonunitary states.
We do not show the derivation for unitary states because it 
is much simpler than that in nonunitary states.
As shown in above equations, 
the expression of the ARC's in nonunitary states is very complicated.
However if a relation
\begin{align}
\boldsymbol{d}=&\boldsymbol{d}_+=\nu \boldsymbol{d}_-,\\
\nu =& 1\,\, \textrm{ or}\,\, -1,
\end{align}
is satisfied, the ARC's can be 
reduced to a rather simple expression 
\begin{align}
{\hat r}_{NN}^{eh}(\boldsymbol{p},L)=& \left.-i
 \left\{ \frac{{\bar k}_z^2}{2|\boldsymbol{q}|} \sum_{l=1}^2
 \frac{ {\hat P}_l}{\Xi_{nu}(l)} 
\right\} {\hat \Delta} \right|_L\!\!\!\!\textrm{ e}^{i\varphi_L},\\
{\hat r}_{NN}^{he}(\boldsymbol{p},R)=& \left.-i {\hat \Delta}^\dagger
 \left\{ \frac{{\bar k}_z^2}{2|\boldsymbol{q}|}\sum_{l=1}^2
 \frac{ {\hat P}_l}{\Xi_{nu}(l)} 
\right\}  \right|_R\!\!\!\textrm{ e}^{-i\varphi_R},\\
\Xi_{nu}(l) =& H^2 \left\{ (1-\nu)|\omega_n|+(1+\nu)\Omega_l\right\}
\nonumber
\\
 &+{\bar k}_z^2 (|\omega_n|+\Omega_l)\label{xinu}.
\end{align}
The effects of the ZES's on the ARC's can be
easily confirmed in Eqs.~(\ref{xisu}), (\ref{xitu}) and (\ref{xinu}). 
For instance in Eq.~(\ref{xinu}), we find in the limit of
$H>>1$ and $\omega_n\to0$,
\begin{equation}
\Xi_{nu}(l)\to \begin{cases}
2H^2|\Delta_l| & (\nu=1)\\
{\bar k}_z^2 |\Delta_l| & (\nu=-1)
\end{cases}.
\end{equation}
In the absence of the ZES's ($\nu =1$), the reflection coefficients 
proportional to $1/H^2$. On the other hand in the presence of the ZES's
($\nu=-1$), the reflection coefficients are independent of the 
barrier height. In this way, the low-temperature anomaly of the Josephson 
current is described by the ARC's.

In the normal metal, the two Green functions in Eqs.~(\ref{te}) and (\ref{th}) 
satisfy a relation as shown in Appendix B,
\begin{equation}
{\cal G}_{\omega_n}^{N,h}(\boldsymbol{r}',\boldsymbol{r})=
-{\hat \sigma}_2 
\left[\hat{\cal G}_{\omega_n}^{N,e}(\boldsymbol{r},\boldsymbol{r}')\right]^\dagger
{\hat \sigma}_2, \label{ghge}
\end{equation}
because of the time reversal symmetry.
The transmission coefficients can be parameterized by
\begin{align}
{\hat \tau}(\boldsymbol{p}',\boldsymbol{p}) \equiv& 
 \tau_0(\boldsymbol{p}',\boldsymbol{p}){\hat \sigma_0} + {\boldsymbol{\tau}}
(\boldsymbol{p}',\boldsymbol{p})\cdot {\hat {\boldsymbol{\sigma}}},\label{taudef}\\
=&\sqrt{v_{\boldsymbol{p}} v_{\boldsymbol{p}'}}
 \int d{\boldsymbol{\rho}} \int d{\boldsymbol{\rho}}'
 \chi_{\boldsymbol{p}'}^\ast({\boldsymbol{\rho}}') 
\chi_{\boldsymbol{p}}({\boldsymbol{\rho}})\nonumber \\
 &\times {\hat {\cal G}}_{\omega_n}^{N,e}
({\boldsymbol{\rho}}',L_N ; {\boldsymbol{\rho}},0).
\end{align}
Since the amplitude of the spin-orbit scattering is much smaller than that
of the spin-independent transmission probability, we assume that
\begin{equation}
|\tau_0|\gg |{\boldsymbol{\tau}}|.
\end{equation}    
The conductance of the normal metal at $T=0$ is given by
\begin{align}
G_N=&\lim_{\omega_n\to 0}\frac{e^2}{h} \textrm{ Tr} 
\sum_{\boldsymbol{p},\boldsymbol{p}'} 
{\hat \tau}(\boldsymbol{p}',\boldsymbol{p})
{\hat \tau}^\dagger(\boldsymbol{p}',\boldsymbol{p}),\\
\simeq&\lim_{\omega_n\to 0}\frac{2e^2}{h} 
\sum_{\boldsymbol{p},\boldsymbol{p}'}
|\tau_0(\boldsymbol{p}',\boldsymbol{p})|^2.
\end{align}

By using Eq.(\ref{taudef}), the Josephson current is rewritten as
\begin{align}
J=&-2e\textrm{ Im} 
\sum_{\boldsymbol{p}} \sum_{\boldsymbol{p}'} 
T\sum_{\omega_n} \textrm{ Tr} \nonumber \\
&\times 
\left[ \hat{r}_{NN}^{eh}(\boldsymbol{p},L) \cdot {\hat \sigma}_2
\{
\tau_0^\ast{\hat \sigma_0} + 
{\boldsymbol{\tau}}^\ast 
\cdot 
{\hat {\boldsymbol{\sigma}}} \} {\hat \sigma}_2 \right.
\nonumber \\
& \cdot \left. \hat{r}_{NN}^{he}(\boldsymbol{p}',R) \cdot 
\{\tau_0{\hat \sigma_0} + {\boldsymbol{\tau}}
\cdot {\hat {\boldsymbol{\sigma}}}\}
\right].
\end{align}
Firstly we consider Josephson junctions where
the two superconductors have the spin-singlet Cooper pairs.
The Josephson current is given by
\begin{align}
J_{SS}=& 4e \sin\varphi T\sum_{\omega_n} \sum_{\boldsymbol{p},\boldsymbol{p}'}
\nonumber\\ 
&\times \Gamma_{su}(\boldsymbol{p}',R) 
|\tau_0(\boldsymbol{p}',\boldsymbol{p})|^2 
\Gamma_{su}(\boldsymbol{p},L),\label{jss}
\end{align}
where $\varphi=\varphi_L-\varphi_R$.

Secondly we consider junctions where spin-triplet 
and spin-singlet superconductors
are on the left and on the right hand sides, respectively.
The Josephson current 
results in
\begin{align}
 J_{TS}=& 4e T\sum_{\omega_n} \sum_{\boldsymbol{p},\boldsymbol{p}'}
\textrm{Im}
\nonumber\\ 
 \times&\left[ \textrm{ e}^{i\varphi}
\Gamma_{su}(\boldsymbol{p}',R)   
\boldsymbol{W}(\boldsymbol{p}',\boldsymbol{p}) \cdot 
{\boldsymbol{\Gamma}}_{t}(\boldsymbol{p},L)
\right],\label{jts}\\
 \boldsymbol{W}(\boldsymbol{p}',\boldsymbol{p})=&
\left( \tau_0^\ast {\boldsymbol{\tau}}
+\tau_0{\boldsymbol{\tau}}^\ast +i {\boldsymbol{\tau}}^\ast\times
{\boldsymbol{\tau}} \right)(\boldsymbol{p}',\boldsymbol{p}),\label{vecw}
\end{align}
where 
${\boldsymbol{\Gamma}}_{t}$ represents ${\boldsymbol{\Gamma}}_{tu}$
in Eq.~(\ref{gammatu})
or ${\boldsymbol{\Gamma}}_{nu}$ in Eq.~(\ref{gammanu}).
As shown in Eq.~(\ref{vecw}), the $J_{TS}$ vanishes in the absence of the 
spin-orbit
scattering in the normal metal~\cite{geshkenbein,millis,sigrist2}.

Finally when the two superconductors have spin-triplet Cooper pairs,
the Josephson current is given by 
\begin{align}
J_{TT}=& 4e T\sum_{\omega_n} \sum_{\boldsymbol{p},\boldsymbol{p}'}\textrm{ Im}
\nonumber\\ 
\times&\left[ \textrm{ e}^{i\varphi}
{\boldsymbol{\Gamma}}_{t}(\boldsymbol{p},L) \cdot  
{\boldsymbol{\Gamma}}_{t}^\ast(\boldsymbol{p}',R) 
|\tau_0(\boldsymbol{p}',\boldsymbol{p})|^2 \right].\label{jtt}
\end{align}

The obtained formula in Eqs.~(\ref{jss}), (\ref{jts}) and (\ref{jtt})
are essentially the same as those in the previous results~\cite{sigrist2}
when the ZES's are not formed at the NS interfaces.
However in the presence of the ZES's, the dependence of the 
Josephson current on temperatures in our results is drastically 
different from that in the previous one's.  
This is because the ARC's ( $\Gamma_{su}$,
${\boldsymbol{\Gamma}}_{tu}$ and ${\boldsymbol{\Gamma}}_{nu}$)
describe the low-temperature anomaly of the Josephson current in the
SNS junctions of anisotropic superconductors.

\section{Conclusion}
On the basis of the mean-field theory of the superconductivity, we 
derive a formula for the Josephson current between two anisotropic
superconductors.
The Josephson current is expressed by the Andreev reflection
coefficients at the junction interfaces.
The contribution of the zero-energy
bound states formed at the NS interfaces to the Josephson current 
is taken into account through these Andreev reflection coefficients. 
The formula can be applied to SIS and
SNS junctions of the anisotropic superconductors 
with spin-singlet and spin-triplet Copper pairs.

\section*{Acknowledgements}
The author is indebted to N.~Tokuda, H.~Akera and Y.~Tanaka
for useful discussion. We also thank N.~Hatakenaka for sending
their preprint.

\appendix
\section{Transmission and Reflection Coefficients at the NS interface}

\begin{figure}[h]
\includegraphics[width=8.0cm]{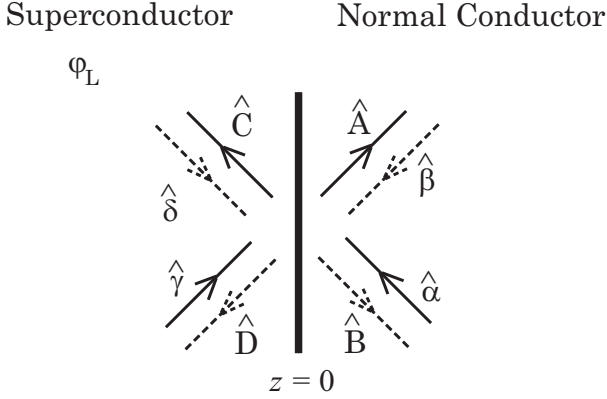}
\caption{Amplitudes of incoming and outgoing waves at the left NS
interface. 
}
\label{app1}
\end{figure}

We derive the transmission and the reflection coefficients
at the left NS interface ($z=0$), where the superconductor is in 
spin-triplet nonunitary states as shown in Fig.~\ref{app1}. 
In what follows, we calculate the coefficients after the analytic
continuation (i.e., $E\rightarrow i\omega_n$) for $\omega_n > 0$. 
In the normal metal, a wavefunction of a quasiparticle 
can be described by
\begin{widetext}
\begin{equation}
{\boldsymbol{\Psi}}^N_{\boldsymbol{p}}(\boldsymbol{\rho}, z) = 
\left[
 \left( \begin{array}{c} \hat{\alpha} \\ \hat{0} \end{array} \right)
\textrm{ e}^{-i k_z z}
+ \left( \begin{array}{c} \hat{0} \\ \hat{\beta} \end{array} \right)
\textrm{ e}^{ i k_z z}
+ 
 \left( \begin{array}{c} \hat{A} \\ \hat{0} \end{array} \right)
\textrm{ e}^{ i k_z z}
+ \left( \begin{array}{c} \hat{0} \\  \hat{B} \end{array} \right)
\textrm{ e}^{-i k_z z}
\right] \chi_{\boldsymbol{p}}({\boldsymbol{\rho}}),
\end{equation}
where $\hat{\alpha}$ and $\hat{\beta}$ ($\hat{A}$ and $\hat{B}$ ) 
are the amplitudes of  
incoming (outgoing) waves
in the electron and the hole branches, respectively.
In the same way, a wavefunction in the superconductor is
given by %
\begin{align}
 {\boldsymbol{\Psi}}^S_{\boldsymbol{p}}(\boldsymbol{\rho},z) 
=& \check{\Phi}_L\left[
\left( \begin{array}{c} \hat{u}_+ \\ \hat{\Delta}_+^\dagger \hat{v}_+ 
\end{array} \right)
\textrm{ e}^{ i k_z z}\hat{\gamma}
+ 
\left( \begin{array}{c} \hat{\Delta}_-\hat{v}_-^\ast 
\\ \hat{u}_-^\ast \end{array} \right)
\textrm{ e}^{-i k_z z}\hat{\delta}
+  
\left( \begin{array}{c} \hat{u}_-  \\ \hat{\Delta}_-^\dagger \hat{v}_- 
\end{array} \right)
\textrm{ e}^{-i k_z z}\hat{C} \right. \nonumber \\
&+    
\left. \left( \begin{array}{c} \hat{\Delta}_+ \hat{v}_+^\ast 
\\ \hat{u}_+^\ast \end{array} 
\right)
\textrm{ e}^{i k_z z}\hat{D}
\right]  \chi_{\boldsymbol{p}}(\boldsymbol{\rho}),
\end{align}
where $\hat{\gamma}$ and $\hat{\delta}$ ($\hat{C}$ and $\hat{D}$ )
are the amplitudes of incoming (outgoing) waves
in the electron and the hole branches, respectively. 
We note that $\hat{\alpha}$, $\hat{\beta}$, $\hat{\gamma}$ and 
$\hat{\delta}$ have only diagonal elements.

The two wavefunctions satisfy a continuity-condition at the left
NS interface,
\begin{align}
{\boldsymbol{\Psi}}^N_{\boldsymbol{p}}(\boldsymbol{\rho},0)=& 
{\boldsymbol{\Psi}}^S_{\boldsymbol{p}}(\boldsymbol{\rho},0), 
\label{conti1}\\
 \left. \frac{\partial}{\partial z} {\boldsymbol{\Psi}}^N_{\boldsymbol{p}}
(\boldsymbol{\rho},z)\right|_{z=0}
- 2mV_b {\boldsymbol{\Psi}}^N_{\boldsymbol{p}}(\boldsymbol{\rho},0) 
=& \left. \frac{\partial}{\partial z} {\boldsymbol{\Psi}}^S_{\boldsymbol{p}}
(\boldsymbol{\rho},z)\right|_{z=0}
\label{conti2}.
\end{align} 
\end{widetext}
From Eqs.~(\ref{conti1}) and (\ref{conti2}), we obtain 
the transmission and the reflection coefficients
\begin{align}
\hat{t}_{SN}^{ee}(\boldsymbol{p},L)=& \bar{k}_z \, \kappa\, \hat{u}_-^{-1} 
\, \hat{Z}_2^\dagger\,  \hat{\xi}_{2,+}^\dagger\, 
\textrm{ e}^{-i\varphi_L/2},\label{snee} \\
\hat{t}_{SN}^{eh}(\boldsymbol{p},L)=& \bar{k}_z \, H \, \hat{u}_-^{-1}\,  
\hat{Z}_2^\dagger \,  \textrm{ e}^{i\varphi_L/2},\\
\hat{t}_{SN}^{he}(\boldsymbol{p},L)=& -\bar{k}_z \, H 
\, \left(\hat{u}_+^\ast \right)^{-1} \,  \hat{Z}_1\, 
\textrm{ e}^{-i\varphi_L/2},\\
\hat{t}_{SN}^{hh}(\boldsymbol{p},L)=& \bar{k}_z \, \kappa^\ast\,  
\left(\hat{u}_+^\ast \right)^{-1} \,  \hat{Z}_1
\, \hat{\xi}_{2,-} \, \textrm{ e}^{i\varphi_L/2},\\
\hat{r}_{SS}^{eh}(\boldsymbol{p},L)=& -\hat{u}_-^{-1} \hat{\Delta}_- \,  
\hat{v}_-^\ast \nonumber \\
 &-i \frac{H^2}{\omega_n} \, \hat{u}_-^{-1} \,  
\hat{Z}_2^\dagger \,
\left(\hat{u}_-^t \right)^{-1} \hat{\Omega}_-,\\
\hat{r}_{SS}^{he}(\boldsymbol{p},L)=& -\left( \hat{u}_+^\ast\right)^{-1} 
\hat{\Delta}_+^\dagger \hat{v}_+ \nonumber \\
& \!\!\!\!\!-i \frac{H^2}{\omega_n} \left(\hat{u}_+^\ast\right)^{-1} 
\hat{Z}_1
\left(\hat{u}_-^\dagger \right)^{-1} \hat{\Omega}_+,
\\
\hat{r}_{NN}^{eh}(\boldsymbol{p},L)=& -i {\bar k}_z^2 \, 
 \hat{Z}_2^\dagger \, 
\textrm{ e}^{i\varphi_L}, \\
\hat{r}_{NN}^{he}(\boldsymbol{p},L)=& -i \bar{k}_z^2 \,
\hat{Z}_1 \, 
\textrm{ e}^{-i\varphi_L}, \\
\hat{t}_{NS}^{ee}(\boldsymbol{p},L)=& \frac{\bar{k}_z \kappa}{\omega_n} \, 
 \hat{Z}_2^\dagger \, \hat{\xi}_{2,+}^\dagger \,  
 \left( \hat{u}_+^\dagger\right)^{-1}\hat{\Omega}_+\textrm{ e}^{i\varphi_L/2},
\end{align}
\begin{align}
\hat{t}_{NS}^{eh}(\boldsymbol{p},L)=& -\frac{\bar{k}_z H}{\omega_n} \, 
\hat{Z}_2^\dagger \, 
 \left( \hat{u}_-^t\right)^{-1}\hat{\Omega}_- \,
\textrm{ e}^{i\varphi_L/2},\\ 
\hat{t}_{NS}^{he}(\boldsymbol{p},L)=& \frac{\bar{k}_z H}{\omega_n}  
\, \hat{Z}_1 
\left(\hat{u}_+^\dagger\right)^{-1} \hat{\Omega}_+
\, \textrm{ e}^{-i\varphi_L/2}\\
\hat{t}_{NS}^{hh}(\boldsymbol{p},L)=& \frac{\bar{k}_z \kappa^\ast}{\omega_n}
\, \hat{Z}_1 \, \hat{\xi}_{2,-}
\left(\hat{u}_+^t\right)^{-1} \hat{\Omega}_- 
\textrm{ e}^{-i\frac{\varphi_L}{2}},\label{nshh}\\
\hat{r}_{NN}^{he}(-\boldsymbol{p},L)=&
\left[\hat{r}_{NN}^{eh}(\boldsymbol{p},L)\right]^\ast.
\end{align}
Here we define
\begin{align}
\hat{\xi}_{1,\pm}=&\left(\frac{1}{2|\boldsymbol{q}_\pm|}\sum_{l=1}^2 
\frac{K_{l,\pm}}{|\Delta_{l,\pm}|^2} \hat{P}_{l,\pm}\right) \hat{\Delta}_\pm,\\
\hat{\xi}_{2,\pm}=&\left(\frac{1}{2|\boldsymbol{q}_\pm|}\sum_{l=1}^2 
\frac{\hat{P}_{l,\pm}}{K_{l,\pm}} \right) \hat{\Delta}_\pm,\\
\hat{Z}_1 =& \left[ H^2 \hat{\xi}_{1,+} +|\kappa|^2 \hat{\xi}_{2,-}
\right]^{-1},\\
\hat{Z}_2 =& \left[ H^2 \hat{\xi}_{1,-} +|\kappa|^2 \hat{\xi}_{2,+}
\right]^{-1},\\
\kappa =& \bar{k}_z +i H.
\end{align}
In the same way, the ARC's 
at the right NS interface are given by
\begin{align}
\hat{r}_{NN}^{he}(\boldsymbol{p},R)=&-i \bar{k}_z^2 \,
\hat{Z}_2 \, \textrm{ e}^{-i\varphi_R},\\
\hat{r}_{NN}^{eh}(-\boldsymbol{p},R)=& 
\left[\hat{r}_{NN}^{he}(\boldsymbol{p},R)\right]^\ast.
\end{align}

On the derivation, we use identities,
\begin{align}
\hat{S}_{l,\pm}^\dagger \cdot \hat{S}_{l',\pm}
=&\frac{\hat{t}_l}{2Q_\pm^2} \, \delta_{l,l'},\\
\hat{S}_{l,\pm} \cdot \hat{S}_{l',\pm}^\dagger 
=& \frac{\hat{P}_{l,\pm}}{2|\boldsymbol{q}_\pm |Q_\pm^2}\, \delta_{l,l'},\\
\hat{P}_{l,\pm} \cdot \hat{P}_{l',\pm}=& 2|\boldsymbol{q}_\pm| 
\hat{P}_{l,\pm} \, \delta_{l,l'},\\
\hat{S}_{l,\pm}^\dagger \cdot \hat{\Delta}_\pm \cdot \hat{\Delta}^\dagger_\pm 
\cdot \hat{S}_{l',\pm} =&
\frac{|\Delta_{l,\pm}|^2}{2Q^2_\pm} \, \hat{t}_l \, \delta_{l,l'},\\
\hat{\Delta}^\dagger_\pm \cdot \hat{S}_{l,\pm} 
\cdot \hat{S}_{l',\pm}^\dagger \cdot 
\hat{\Delta}_\pm=&
\frac{|\Delta_{l,\pm}|^2}{2|\boldsymbol{q}_\pm|Q^2_\pm} \, 
\hat{P}_{l,\pm}^\ast \, \delta_{l,l'},\\
\hat{\Delta}_{\pm} \cdot \hat{P}_{l,\pm}^\ast =& 
\hat{P}_{l,\pm} \cdot \hat{\Delta}_\pm,\\
\hat{P}_{l,\pm} \cdot \hat{\Delta}_\pm \cdot \hat{\Delta}^\dagger_\pm 
=& |\Delta_{l,\pm}|^2 \hat{P}_{l,\pm}.
\end{align}
The ARC's of superconductors in unitary
states can be calculated in the same way. The derivation  
of the ARC's in unitary states 
is much simpler than that in nonunitary states. 

\begin{figure}[bp]
\includegraphics[width=8.0cm]{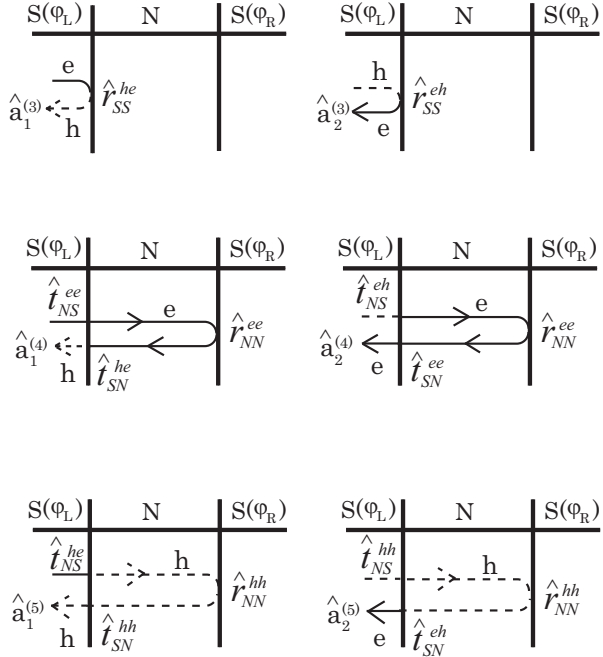}
\caption{Reflection processes included in the coefficients 
$\hat{a}_1$ and $\hat{a}_2$. 
These processes, however, do not contribute to the Josephson current.} 
\label{app2}
\end{figure}

In addition to the four reflection processes shown in Fig.~\ref{process}(a),
six reflection processes can be considered for $\hat{a}_1$ and $\hat{a}_2$ 
as shown in Fig.~\ref{app2}.
By using the coefficients in Eqs.~(\ref{snee})-(\ref{nshh}), 
it is possible to show that these six processes do not 
contribute to the Josephson current.

\section{Transmission coefficients in normal metal}
Since the amplitude of the pair potential 
in the normal metal is taken to be zero, the BdG equation in Eq.(\ref{bdg})
is decoupled into two equations,
\begin{align}
\hat{h}_0(\boldsymbol{r}) \hat{u}_\lambda =& \hat{u}_\lambda \hat{E}_\lambda 
\label{nbdg1}\\
- \hat{h}_0^\ast(\boldsymbol{r}) \hat{v}_\lambda
=& \hat{v}_\lambda
\hat{E}_\lambda.\label{nbdg2}
\end{align}
The Green function in the normal metal obeys the equation,
\begin{align}
(i\omega_n \hat{\sigma}_0 -\hat{h}_0 (\boldsymbol{r}) )
\hat{\cal G}_{\omega_n}^{N,e}
(\boldsymbol{r},\boldsymbol{r}') =& 
\delta(\boldsymbol{r}-\boldsymbol{r}')\hat{\sigma}_0,\\ 
(i\omega_n \hat{\sigma}_0 + \hat{h}_0^\ast (\boldsymbol{r}) )
\hat{\cal G}_{\omega_n}^{N,h}
(\boldsymbol{r},\boldsymbol{r}') =& 
\delta(\boldsymbol{r}-\boldsymbol{r}')\hat{\sigma}_0. 
\end{align}
The Green function in the two branch are represented by
\begin{align}
\hat{\cal G}_{\omega_n}^{N,e}(\boldsymbol{r},\boldsymbol{r}')=&
\sum_{\lambda}
\hat{u}_\lambda(\boldsymbol{r}) 
\left[ i\omega_n\hat{\sigma}_0 - \hat{E}_\lambda
\right]^{-1} \!\!\!\hat{u}_\lambda^\dagger(\boldsymbol{r}'),\\
\hat{\cal G}_{\omega_n}^{N,h}(\boldsymbol{r},\boldsymbol{r}')=&
-\left[\hat{\cal G}_{\omega_n}^{N,e}(\boldsymbol{r},\boldsymbol{r}')
\right]^\ast,
\end{align}
where we use the complex conjugate of Eq.~(\ref{nbdg1}) for the 
Green function in the hole branch. 
By using Eqs.(\ref{te}) and (\ref{th}), we can show relations 
\begin{align}
\hat{t}_{-\boldsymbol{p},-\boldsymbol{p}'}^e =& 
\left[ \hat{t}_{\boldsymbol{p},\boldsymbol{p}'}^h\right]^\ast,\\
\hat{t}_{-\boldsymbol{p}',-\boldsymbol{p}}^h =& 
\left[ \hat{t}_{\boldsymbol{p}',\boldsymbol{p}}^e\right]^\ast.
\end{align}

When the time-reversal symmetry holds in the normal metal,
we find
\begin{equation}
\hat{h}_0^\ast(\boldsymbol{r}) i\hat{\sigma}_2 \hat{u}_\lambda 
= i \hat{\sigma}_2 \hat{u}_\lambda \hat{E}_\lambda.
\end{equation}
The Green function in the hole branch is described by that in the
electron branch,
\begin{equation}
\hat{\cal G}_{\omega_n}^{N,h}(\boldsymbol{r}',\boldsymbol{r}) =-\hat{\sigma}_2
\left[\hat{\cal G}_{\omega_n}^{N,e}(\boldsymbol{r},\boldsymbol{r}')\right]^\dagger
\hat{\sigma}_2.
\end{equation}

\end{document}